\documentstyle[preprint,tighten,aps]{revtex}
\begin{document} 
\draft
\preprint{IASSNS-HEP-97/62,hep-th/9706046}
\date{March 1997}
\title{Entropy of Non-Extreme Rotating Black Holes in String Theories}
\author{Donam Youm
\thanks{E-mail address: youm@sns.ias.edu}}
\address{School of Natural Sciences, Institute for Advanced
Study\\ Olden Lane, Princeton, NJ 08540}
\maketitle
\begin{abstract}
{We formulate the Rindler space description of rotating black 
holes in string theories.  We argue that the comoving frame is 
the natural frame for studying thermodynamics of rotating black 
holes and statistical analysis of rotating black holes gets simplified 
in this frame.  We also calculate statistical entropies of general 
class of rotating black holes in heterotic strings on tori by applying 
$D$-brane description and the correspondence principle.  We find at least 
qualitative agreement between the Bekenstein-Hawking entropies and 
the statistical entropies of these black hole solutions.}
\end{abstract}

\section{Introduction}\label{intro}

Past couple of years have been revolutionary period for understanding 
one of challenging problems in quantum theory of gravity.  Namely, 
we are now able to reproduce Bekenstein-Hawking entropy 
\cite{BEK73,BEK74,HAW71,HAW13,BARch31} of special 
class of black holes through stringy statistical calculations without 
encountering infinities associated with non-renomalization of 
quantum gravity or information loss problems \cite{HOO256}. 
 
This was first anticipated in Ref.\cite{SUS45}, where it is predicted that 
since string theory is a finite theory of quantum gravity the calculation 
of partition function in the canonical quantum gravity of superstring 
theories would yield finite statistical entropy of black holes.  
In this description, the microscopic degenerate degrees of freedom 
responsible for non-zero statistical entropy of black holes are 
oscillating degrees of freedom of strings which are in thermal 
equilibrium with black hole environment \cite{SUSu50}.  
(Only strings that contribute to statistical entropy are those that 
are somehow entangled with the event horizon and therefore look like 
oscillating open strings whose ends are attached to the event horizon 
from the point of view of external observers.)   

The explicit calculation of the degeneracy of (microscopic) 
string states that are associated with (macroscopic) black hole became 
possible with identification \cite{DUFr} of subsets of black hole 
solutions in effective string theories with massive string states and 
the level matching condition \cite{DABghw474}, which justifies such  
identification.  The qualitative agreement (up to a factor of numerical 
constant of order 1) was first observed in Ref.\cite{SEN10}, where the 
Bekenstein-Hawking entropy of NS-NS electric charged, non-rotating, 
black holes in heterotic string which is calculated at the stretched 
horizon is compared with degeneracy of massive perturbative string states.  

Precise calculation of statistical entropy was made possible with 
construction of general class of black hole solutions 
(for example those constructed in Refs.\cite{CVEy672,CVEy476}) which 
have non-zero event horizon area in the BPS limit.  
Such solutions necessarily carry non-perturbative charges such as 
magnetic charges or charges in the R-R sector of string theories.  
The first attempt to calculate string state degeneracy of the corresponding 
black holes that carry non-perturbative charges is based upon 
throat limit of a special class of conformal model, called chiral 
null model \cite{CVEt53,TSE11,TSE477}.  
In this description, the throat region is described by 
a WZNW model with the level determined by NS-NS magnetic charges 
(carried by black hole solutions).  Effectively, the throat region 
conformal model describing the general class of dyonic solution is 
the sigma model for perturbative heterotic string with string tension 
rescaled by magnetic charges.  

Another stringy description of dyonic black holes, which is accepted widely 
and can be applicable to non-BPS black holes as well, was prompted by the 
realization \cite{POL75} that non-perturbative R-R charges of type-II 
string theories can be carried by $D$-branes \cite{DAIlp4}, i.e. boundaries 
upon which open strings with Dirichlet boundary conditions are constrained 
to live.  
With introduction of $D$-branes into string states as carriers of 
non-perturbative R-R charges, it becomes possible to circumvent 
\cite{SEN54,SEN53,VAF088,VAF463,BERsv398,BERsv463} the previous 
difficulty of having to count bound states of solitons and perturbative 
string states when one has to determine the degeneracy of non-perturbative 
string states that carry R-R charges as well as NS-NS electric charges.  
In the $D$-brane description of black holes \cite{STRv379,CALm472}, 
microscopic degrees of freedom of black holes originate from oscillating 
degrees of freedom of open strings which stretch between $D$-branes.  
Effectively, $D$-brane bound state description of black holes is 
that of perturbative open strings with central charge rescaled by 
R-R charges or the number of $D$-branes.  

In the chiral null model and the $D$-brane descriptions of dyonic 
black holes, carriers of non-perturbative charges play non-dynamic 
roles gravitationally and act just as backgrounds in which strings 
oscillate, effectively playing the role of rescaling string tensions 
or central charges.  (In principle, massive string excitations will 
modify gravitational field of black holes.)  
Base upon this observation and 't Hooft's idea \cite{HOO256} that 
entropy of black holes is nothing but the entropy of fields in thermal 
equilibrium with black hole environment, Susskind and others 
\cite{HALrs392,HALkrs75,HAL68,HAL75} proposed braneless description 
of black holes, in an attempt to solve problems associated with 
stringy statistical interpretation of non-extreme black holes 
\cite{SUS45}.  
Namely, in their description of `braneless' black holes, 
non-perturbative charges are carried by 
black holes which act as background gravitational field in which 
perturbative, NS-NS electric charged strings oscillate.  
Due to scaling of the time coordinate to the Rindler time nearby 
the event horizon, string tension and string oscillation levels 
get scaled
\footnote{Note, the time coordinate is conjugate to energy, since 
Hamiltonian is the generator for time-translation.} 
by non-perturbative charges.  Using this prescription, they were able 
to reproduce Bekenstein-Hawking entropies of non-rotating black holes 
in five [four] dimensions with three [four] charges in the BPS and 
near BPS limit \cite{HAL68,HAL75}.  

It is one of purposes of this paper to generalize 
their argument to the case of rotating black holes in string theories.  
In the comoving coordinates, geometry of rotating black holes 
approximates to that of Rindler spacetime in the region close to the 
event horizon, i.e. the throat region.  Thereby, the sigma-model 
description of (rotating) black hole background gets simplified in 
the comoving frame.  We shall build up frameworks for understanding 
statistical description of rotating black holes in the Rindler spacetime 
picture and speculate on some of points that we do not have complete 
understanding of.  We believe that precise agreement of the 
Bekenstein-Hawking entropy and the statistical entropy in this 
description will require better understanding of subtleties on 
perturbative strings in the comoving frame.  

Main justification for (perturbative) stringy calculation of 
statistical entropy of black holes hinges on a special property 
of BPS states that the degeneracy of BPS states is a topological 
quantity which is independent of coupling constants.  Thereby, one 
can safely calculate the microscopic degeneracy of black holes, which 
correspond to strong coupling limit, in the weak coupling limit, 
in which spacetime is Minkowskian and perturbative description of 
strings is valid.  Also, it is supersymmetry (preserved by BPS 
solutions) that renders quantum corrections under control.  
On the other hand, it is observed (to the contrary to conventional 
lore) that even for non-BPS extreme solutions \cite{MALs77,DAB050}, 
near extreme solutions \cite{HORs77,BRElmpsv381,HORlm77,HORms383,MAL125} 
and non-extreme solutions \cite{HORp46}, $D$-brane description of 
black holes reproduces Bekenstein-Hawking entropy 
correctly.  In particular, according to the correspondence principle 
proposed in Ref.\cite{HORp46}, the Bekenstein-Hawking entropies of  
non-extreme black holes can be reproduced by $D$-brane or perturbative 
string descriptions when the size of the event horizon is of the 
order of string scale.  The correspondence principle gives rise to 
statistical entropy of non-rotating, non-extreme black holes which 
is qualitatively in agreement with the Bekenstein-Hawking entropy 
up to a numerical factor of order one.  In this paper, we shall 
show that the correspondence principle can be applied to special  
class of electrically charge non-extreme, rotating black holes in 
heterotic string on tori, as well.  

The paper is separated into two parts.  In the first part, we shall 
attempt to formulate Rindler space description of rotating black holes.  
In the second part, we discuss statistical interpretation of general 
class of rotating black holes in heterotic string on tori 
that are constructed in Refs.\cite{CVEy476,CVEy54,CVEy477} by applying 
$D$-brane picture and the correspondence principle.  
In section \ref{rotbhst}, we shall discuss some of global spacetime properties 
of rotating black holes and derive that in the comoving frame the spacetime 
approximates to the Rindler spacetime in the region very close to the event 
horizon.  In section \ref{statcmf}, we discuss statistical description 
of strings in rotating black hole background and in particular we show that 
the entropy of strings in the rotating frame is the same as the 
entropy in the static frame.  In section \ref{bhsol}, we summarize properties 
of general classes of non-extreme, rotating black holes in 
heterotic string on tori constructed in Refs.\cite{CVEy476,CVEy54,CVEy477}.  
In section \ref{statent}, we discuss the statistical description of such 
black hole solutions in the picture of $D$-brane and the correspondence 
principle, and speculate on the Rindler spacetime description of statistical 
entropy.

\section{Spacetime Properties of Rotating Black Holes}\label{rotbhst}

In this section we summarize some of properties of rotating black holes 
necessary in understanding thermodynamics of fields that are 
in thermal equilibrium with the black hole environment.  
In addition to the ADM mass and $U(1)$ charges, rotating black holes 
are characterized by angular momenta.  The presence of angular momenta  
dramatically changes the global spacetime properties of black holes.  

In general, spacetime metric for axisymmetric black holes in $D$-dimensional 
spacetime can be written in the following form in the Boyer-Lindquist 
coordinates:
\begin{equation}
ds^2=g_{tt}dt^2+g_{rr}dr^2+2g_{t\phi_i}dtd\phi_i+
g_{\phi_i\phi_j}d\phi_id\phi_j+g_{\theta\theta}d\theta^2
+2g_{\theta\psi_i}d\theta d\psi_i+g_{\psi_i\psi_j}d\psi_id\psi_j,
\label{genaximetantz}
\end{equation}
where $\phi_i$ ($i=1,...,[{{D-1}\over 2}]$) correspond to angular 
coordinates in the $[{{D-1}\over 2}]$ orthogonal rotational planes 
and the index $i$ in other angular coordinates $\psi_i$ runs from 
1 to $[{{D-4}\over 2}]$.  The metric coefficients $g_{\mu\nu}$ 
($\mu,\nu=0,1,...,D-1$) in the Boyer-Lindquist coordinates are 
independent of $t$ and $\phi_i$, manifesting ``time-translation 
invariant'' (stationary) and ``axially symmetric'' spacetime geometry.  
So, the Killing vectors associated with these symmetries are 
$\xi_{(t)}\equiv\partial/\partial_t$ and $\xi_{(\phi_i)}\equiv
\partial/\partial_{\phi_i}$.  

Due to axially symmetric spacetime, any observer who moves along the 
worldline of constant $(r,\theta,\psi_i)$ with uniform angular 
velocity does not notice any change in spacetime geometry.  
Hence, such an observer can be thought of as ``stationary'' relative 
to the local geometry.  On the other hand, an observer who moves 
along the constant $(r,\theta,\psi_i,\phi_i)$ worldline, i.e. with zero 
angular velocity, is also ``static'' relative to the asymptotic 
spacetime.  Here, the angular velocity $\Omega_i$ ($i=1,...,
[{{D-1}\over 2}]$) relative to asymptotic rest frame is defined as
$\Omega_i\equiv{{d\phi_i}\over{dt}}$.  
Since an observer cannot move faster than the speed of light, 
the angular velocity $\Omega_i$ of the observer is constrained to 
take limited values.  Namely, the constraint that the $D$-velocity 
${\bf u}=u^t(\partial/\partial t+\Omega_i\partial/\partial\phi_i)=
{{\xi_{(t)}+\Omega_i\xi_{\phi_i}}\over{|\xi_{(t)}+\Omega_i
\xi_{\phi_i}|}}$ ($u^t\equiv{{dt}\over{d\tau}}$) lies inside 
the future light cone, i.e. ${\bf u}\cdot{\bf u}<0$, restricts 
the angular velocity of stationary observers to be bound by 
their minimum and maximum values, i.e. $\Omega_{i\,min}<\Omega_i
<\Omega_{i\,max}$.  The minimum and the maximum values of the 
angular velocity are explicitly in the following forms:
\begin{equation}
\Omega_{i\,min}=\omega_i-\sqrt{\omega^2_i-g_{tt}/g_{\phi_i\phi_i}}, 
\ \ \ \ \ \ 
\Omega_{i\,max}=\omega_i+\sqrt{\omega^2_i-g_{tt}/g_{\phi_i\phi_i}},
\label{minmaxangvel}
\end{equation}
where $\omega\equiv{1\over 2}(\Omega_{i\,min}+\Omega_{i\,max})=
-g_{t\phi_i}/g_{\phi_i\phi_i}$.  
Note, $r\Omega_{i\,min}=-1$ and $r\Omega_{i\,max}=1$ 
at spatial infinity, corresponding to the speed of light in the 
Minkowski spacetime.  

As observers approach the black hole, $\Omega_{i\,min}$ increases 
and finally becomes zero when $g_{tt}=0$.  Therefore, in the region 
at and inside of the hypersurface defined by $g_{tt}=0$, observers 
have to rotate (with positive angular velocities), i.e. 
observers cannot be static relative to the asymptotic rest frame.  
For this reason, this hypersurface defined by $g_{tt}=0$, i.e. 
defined as the surface on which the Killing vector $\xi_t=\partial/
\partial_t$ vanishes, is called ``static limit''.   
As observers further approach the event 
horizon (defined as $g^{rr}=0$ surface, on which $r=constant$ surface 
is null), the range of values that $\Omega_i$ can take on narrows down,  
and finally at the horizon the minimum and maximum values of $\Omega_i$ 
coalesce.  This can be seen by the fact that the value of $\omega_i=
-g_{t\phi_i}/g_{\phi_i\phi_i}$ at the event horizon corresponds to the 
angular velocity $\Omega_{H\,i}$ of the event horizon, which is defined 
by the condition that the Killing vector $\xi\equiv\partial/\partial t +
\Omega_{H\,i}\partial/\partial\phi_i$ is null on the horizon.  
So, at the event horizon, observers cannot be stationary and fall into 
the black hole through the horizon.  Therefore, in the region between 
the static limit and the event horizon (called ``ergosphere'')  
observers are forced to rotate in the direction of the black hole 
rotation (``dragging of inertial frames'').  

Due to the ``dragging of inertial frames'' by the black hole's angular 
momenta, particles or strings which are in equilibrium with the 
thermal bath of black hole nearby the event horizon have to rotate 
with the black hole.  
Therefore, in order to study thermodynamics of fields in thermal 
equilibrium with the black hole environment, one has to understand 
thermodynamics of particles (strings) in the comoving frame which 
rotates with the black hole.  
As we will discuss in the following subsection, in the comoving 
frame which rotates with the angular velocity of the event horizon 
the spacetime metric in the throat region simplifies to 
that of the Rindler spacetime.  In such frame, analysis of 
statistical mechanics of strings gets simpler and one can just 
apply the result of flat-spacetime, non-interacting, perturbative 
string theory for studying statistical entropy of rotating black 
holes.

\subsection{Comoving Frame and Rindler Geometry}\label{cmfrg}

We consider the comoving frame which rotates with an angular 
velocity $\Omega_{i}$ ($i=1,...,[{{D-1}\over 2}]$) by performing the 
coordinate transformation $\phi_i\to\phi^{\prime}_i=\phi_i-\Omega_it$.  
In this comoving frame, the spacetime metric (\ref{genaximetantz}) 
takes the following form:
\begin{eqnarray}
ds^2=g^{\prime}_{\mu\nu}dx^{\prime\,\mu}dx^{\prime\,\nu}&=&
(g_{tt}+2\Omega_ig_{t\phi_i}+\Omega_i\Omega_jg_{\phi_i\phi_j})
dt^2+2(g_{t\phi_i}+\Omega_jg_{\phi_j\phi_i})dtd\phi^{\prime}_i
+g_{\phi_i\phi_j}d\phi^{\prime}_id\phi^{\prime}_j
\cr
& &+g_{rr}dr^2+g_{\theta\theta}d\theta^2+2g_{\theta\psi_i}d\theta 
d\psi_i+g_{\psi_i\psi_j}d\psi_id\psi_j.
\label{comovframmet}
\end{eqnarray}
In the comoving frame, one has to restrict the system of particles 
(strings) to be in the region such that $g^{\prime}_{tt}=
g_{tt}+2\Omega_ig_{t\phi_i}+\Omega_i\Omega_jg_{\phi_i\phi_j}<0$.  
Namely, in the region defined by $g^{\prime}_{tt}>0$, the comoving 
observer has to move faster than the speed of light.  At the boundary  
surface $g^{\prime}_{tt}=0$, the comoving observer has to move 
with the speed of light.  As we shall see in the following section, 
the Helmholtz free energy $F$ diverges in the region $g^{\prime}_{tt}
\geq 0$, thereby the spacetime region under consideration has to 
be restricted to $g^{\prime}_{tt}<0$ so that thermodynamic observables 
are well-defined.  

From now on, we consider the Hartle-Hawking vacuum state, which is 
defined as the system which rotates with the angular velocity of 
the event horizon and therefore (as we will see) the temperature of 
the system equals to the Hawking temperature.  We are interested 
in the throat region ($r\simeq r_H$) of the comoving frame 
of this system.  
For this purpose, we consider only the time $t$ and the radial $r$ parts 
of the spacetime metric, i.e. $ds^2_T=g^{\prime}_{tt}dt^2+
g^{\prime}_{rr}dr^2$.  And we take $\Omega_i$ in Eq.(\ref{comovframmet}) 
to be angular velocity of rotating black hole at the outer horizon, 
i.e. $\Omega_i=\Omega_{H\,i}$.  

Then, the $(t,t)$-component $g^{\prime}_{tt}$ of the metric 
(\ref{comovframmet}) can be expressed in the following suggestive 
form in terms of the null Killing vector $\xi={{\partial}\over{\partial t}}
+\Omega_{H\,i}{{\partial}\over{\partial\phi_i}}$:
\begin{equation}
g^{\prime}_{tt}=g_{tt}+2\Omega_{H\,i}g_{t\phi_i}+\Omega_{H\,i}
\Omega_{H\,j}g_{\phi_i\phi_j}=g_{\mu\nu}\xi^{\mu}\xi^{\nu}\equiv 
-\lambda^2.
\label{ttcompmet}
\end{equation}
Next, we define new radial coordinate $\rho$ in the following way:
\begin{equation}
d\rho^2\equiv g_{rr}dr^2 \Rightarrow \rho=\int^r_{r_H}
\sqrt{g_{rr}}dr.
\label{newradcoord}
\end{equation}
Then, the $(t,t)$-component of the metric (\ref{comovframmet}) 
gets simplified in the near-horizon region as follows:
\begin{eqnarray}
g^{\prime}_{tt}dt^2&=&\xi_{\mu}\xi^{\mu}dt^2=
-(\sqrt{-\xi_{\mu}\xi^{\mu}})^2dt^2
\cr
&=&-\left[\left.{{d\sqrt{-\xi_{\mu}\xi^{\mu}}}\over{d\rho}}
\right|_{r=r_H}\rho\right]^2dt^2=-{1\over{g_{rr}}}
\left[\left.{{d\sqrt{-\xi_{\mu}\xi^{\mu}}}\over{dr}}
\right|_{r=r_H}\right]^2\rho^2dt^2.
\label{ttcompsymplif}
\end{eqnarray}
Note, the surface gravity $\kappa$ at the (outer) event horizon 
$r=r_H$ is defined in terms of the norm $\lambda$ of the null Killing 
vector $\xi$ as $\kappa^2\equiv {\rm lim}_{r\to r_H}\nabla_{\mu}
\lambda\nabla^{\mu}\lambda$.  Therefore, the time-radial parts $ds^2_T$ 
of the metric in the vicinity of the event horizon $r=r_H$ take the 
following simplified form
\footnote{In can be shown that the other comoving frame metric components 
$g^{\prime}_{\mu\nu}$ also get simplified in the throat region.}  
\begin{equation}
ds^2_T=-\kappa^2\rho^2dt^2+d\rho^2.
\label{rindlermet}
\end{equation}
This becomes the Rindler spacetime metric after one rescales 
the time coordinate $t$ to the Rindler time $\tau\equiv\kappa t$ 
and performs Euclidean time coordinate transformation.  
Therefore, we come to the conclusion that 
{\it in the comoving frame of the Hartle-Hawking vacuum system 
the throat region approximates to the Rindler spacetime}.

\section{Statistical Mechanics and Comoving Frame}\label{statcmf} 

We consider the canonical ensemble of statistical system. 
In this description, the statistical system under 
consideration is regarded as a macroscopic body which is 
in thermal equilibrium with some larger ``medium'' of closed 
thermal system with a fixed temperature $T$.  The larger 
system behaves like a heat reservoir.  The statistical 
distribution of the canonical ensemble is given by the Gibbs 
distribution.  Therefore, the canonical ensemble is suitable for 
describing thermodynamics of particles or strings which 
are in thermal equilibrium with the black hole environment.  
In the following, we discuss some basic well-known aspects of 
canonical ensemble for the purpose of setting up notations and 
discussing statistical mechanics in the comoving frame in general 
settings (not restricted to just some special class of rotating 
black holes.)  

The basic quantity of the canonical ensemble is the canonical 
partition function which is defined as a functions of a variable $\beta$:
\begin{equation}
Z(\beta)\equiv \sum_{\alpha}e^{-\beta E_{\alpha}}
={\rm Tr}\,{\rm exp}(-\beta H),
\label{partition}
\end{equation}
where the trace is over the states $\alpha$ with the energy 
$E_{\alpha}$ and $H$ denotes the Hamiltonian of the smaller subsystem.  
Classically, the partition function $Z(\beta)$ corresponds to 
the volume in the phase space occupied by the canonical ensemble.  
Here, $\beta$ is a real constant interpreted as the inverse 
temperature of the larger system with which the statistical 
ensemble under consideration is in thermal equilibrium.  

One can re-express the partition function (\ref{partition}) in the 
following way:
\begin{eqnarray}
Z(\beta)&=&\sum_{\alpha}e^{-\beta E_{\alpha}}
=\sum_{\alpha}\int^{\infty}_0dE\delta(E-E_{\alpha})e^{-\beta E}
\cr
&=&\int^{\infty}_0dE\left[\sum_{\alpha}\delta(E-E_{\alpha})\right]
e^{-\beta E}=\int^{\infty}_0dE\,g(E)e^{-\beta E},
\label{tranpartition}
\end{eqnarray}
where $g(E)=\sum_{\alpha}\delta(E-E_{\alpha})$ is the total density 
of states, i.e. $g(E)dE$ is the total number of energy eigenstates 
$\alpha$ of the system in the energy range from $E$ to $E+dE$.  
Note, from Eq.(\ref{tranpartition}) one can see that the partition function 
$Z(\beta)$ is nothing but the Laplace transform of the total density 
of states $g(E)$.  So, alternatively one can obtain $g(E)$ from 
the partition function through the inverse Laplace transformation 
$g(E)=\int^{\varepsilon+i\infty}_{\varepsilon-i\infty}{{d\beta}\over
{2\pi i}}e^{\beta E}Z(\beta)$, where the contour (defined by a real number 
$\varepsilon$) is chosen to be to the right of all the singularities 
of $Z(\beta)$ in the complex $\beta$ plane.  

The partition function $Z(\beta)$ is related to the Helmholtz 
free energy $F$ in the following way:
\begin{equation}
F=-{1\over\beta}{\rm ln}Z=-T\,{\rm ln}\sum_{\alpha}e^{-E_{\alpha}/T}, 
\label{helmholz}
\end{equation}
where $\beta=1/T$.  
This follows from the following definition of the entropy $S$:
\begin{equation}
S\equiv -\sum_{\alpha}\rho_{\alpha}\ln\rho_{\alpha}=
-{\rm Tr}\,(\rho{\rm ln}\rho),
\label{entdef}
\end{equation}
where $\rho_{\alpha}=e^{-E_{\alpha}/T}/(\sum_{\alpha}e^{-E_{\alpha}/T})$ 
is the ``Gibbs distribution'' or ``canonical distribution'', 
and from the definition of the free energy $F\equiv E-TS$.  
(Here, the energy $E$ is taken as the mean energy $\bar{E}\equiv 
\sum_{\alpha}E_{\alpha}\rho_{\alpha}$.)  So, alternatively entropy $S$ 
can be obtained from the Helmholtz free energy $F$ as
\begin{equation}
S=\beta^2{{\partial F}\over{\partial\beta}}=
-{{\partial F}\over{\partial T}}.
\label{entfree}
\end{equation}
In terms of the total energy density of the state $g(E)$, 
the free energy takes the following form:
\begin{equation}
F={1\over\beta}\int^{\infty}_0dE\,g(E)\,{\rm ln}
\left(1-e^{-\beta E}\right).
\label{freetotden}
\end{equation}

For the canonical ensemble of non-interacting particles (strings) 
that rotate with constant azimuthal angular velocity  
$\Omega_i$, the partition function $Z(\beta)$ in Eq.(\ref{partition}) 
gets modified to
\begin{equation}
Z=\sum_{\alpha}e^{-\beta(E_{\alpha}-\Omega_iJ_{i\,\alpha})}, 
\label{rotpart}
\end{equation}
where $E_{\alpha}$ is the energy in the rest frame and 
$J_i$ are the $\phi_i$-components of angular momentum of the 
particles (strings).  This is a special example of 
the case in which the statistical system has the set of conserved 
quantities $Q$ other than energy $E$.  Examples on the conserved 
quantities $Q$ are angular momentum, string winding number, 
electric charges, etc..  (For this case, the conserved 
quantity $Q$ corresponds to the angular momentum $\vec{J}$.)  
In general, the total energy density of the system with set $Q$ of 
conserved quantities is given by
\begin{equation}
g(E,Q)=\sum_{\alpha}\delta(E-E_{\alpha})\delta_{Q,Q_{\alpha}}.
\label{endenwithcons}
\end{equation}
So, for example, the partition function (\ref{rotpart}) can be derived 
from the total energy density given by (\ref{endenwithcons}) with 
$Q=\vec{J}$ by following the similar procedure as 
Eq.(\ref{tranpartition}).  The final expression is of the form:
\begin{equation}
Z(\beta)=\sum_{J_i}\int^{\infty}_0dE\,g(E,J_i)e^{-\beta(E-\Omega_iJ_i)},
\label{rotpartint}
\end{equation}
where the sum is over the azimuthal quantum numbers of the angular 
momentum $\vec{J}$.  
And the Helmholtz free energy is, therefore, given by
\begin{equation}
F={1\over\beta}\sum_{J_i}\int^{\infty}_0dE\,g(E,J_i)\,{\rm ln}
\left(1-e^{-\beta(E-\Omega_iJ_i)}\right).
\label{rotfree}
\end{equation}

In the following, we show that entropy of rotating systems  
takes the same form whether it is calculated in the comoving-frame 
or in the rest-frame.  Therefore, one can conveniently calculate the 
entropy in the comoving frame of the Hartle-Hawking vacuum system, 
in which the spacetime metric takes the simple Rindler spacetime 
form.  This is consistent with the fact that the entropy 
measures the degeneracy of the number of indistinguishable 
microscopic states, which should be independent of the 
coordinate frame of an observer.   

We consider particles (strings) that rotate with angular velocity 
$\Omega_i$.  In the rest frame, the partition function for such 
rotating canonical ensemble is given by Eq.(\ref{rotpart}) or 
alternatively, in terms of the total energy density, by 
Eq.(\ref{rotpartint}), and the Helmholtz free energy is given by 
Eq.(\ref{rotfree}).    
On the other hand, in the comoving frame (that rotates with the particles or 
the strings) the angular velocity of the system of particles or strings 
is zero.  So, the partition function is of the following form
\begin{equation}
Z(\beta)=\sum_{\alpha}e^{-\beta E^{\prime}_{\alpha}}=
\int^{\infty}_0dE^{\prime}g(E^{\prime})e^{-\beta E^{\prime}},
\label{comovparttran}
\end{equation}
where $E^{\prime}$ is the energy in the comoving frame.  
And correspondingly the Helmholtz free energy is given by
\begin{equation}
F={1\over\beta}\int^{\infty}_0dE^{\prime}g(E^{\prime})\,
{\rm ln}\left(1-e^{-\beta E^{\prime}}\right).
\label{comovfree}
\end{equation}

Note, in the rest frame the requirement that the free energy $F$ in 
Eq.(\ref{rotfree}) takes finite real value restricts the interval of 
the integration over $E$ to be $E>\Omega_iJ_i$.  Namely, when 
$E=\Omega_iJ_i$ the integrand of Eq.(\ref{rotfree}) diverges and 
when $0<E<\Omega_iJ_i$ the integrand takes a complex value.  
For the case of particles (strings) in the background of rotating 
black holes, such restriction of the integration interval 
corresponds to the condition that the speed of particles (strings) 
has to be less than the speed of light, i.e. 
$g^{\prime}_{tt}=g_{tt}+2\Omega_ig_{t\phi_i}+\Omega_i\Omega_j
g_{\phi_i\phi_j}<0$.  Therefore, the Helmholtz free energy $F$ 
in Eq.(\ref{rotfree}) becomes
\begin{eqnarray}
F&=&{1\over\beta}\sum_{J_i}\int^{\infty}_{\Omega_iJ_i}dE\,
g(E,J_i)\,{\rm ln}\left(1-e^{-\beta(E-\Omega_iJ_i)}\right)
\cr
&=&{1\over\beta}\int^{\infty}_0dE\,\sum_{J_i}
g(E+\Omega_iJ_i,J_i)\,{\rm ln}\left(1-e^{-\beta E}\right)
\cr
&=&-\int^{\infty}_0dE\,{1\over{e^{\beta E}-1}}\sum_{J_i}
\Gamma(E+\Omega_iJ_i,J_i),
\label{chgvarfree}
\end{eqnarray}
where in the second equality the change of integration variable 
$E\to E-\Omega_iJ_i$ is performed and in the third equality we have 
integrated by parts.  Here, $\Gamma(E,J_i)$ is the total number of 
states with energy less than $E$, given that the angular momentum 
is $J_i$.  It can be shown that $\sum_{J_i}\Gamma(E+\Omega_iJ_i,J_i)
=\Gamma(E)$.  Therefore, the Helmholtz free energy 
(\ref{rotfree}) in the rest-frame takes the same form as the free energy 
(\ref{comovfree}) in the comoving frame.   Similarly, one can show 
that the partition functions $Z(\beta)$ in the rest-frame and 
the comoving frame have the same form.  Consequently, entropy 
$S=\beta^2{{\partial F}\over{\partial\beta}}$ takes the same values 
in the comoving and the rest frames.  

\subsection{Thermodynamics of the Rindler Spacetime}\label{thrindsp}

We showed in section \ref{cmfrg} that in the comoving frame which rotates 
with the angular velocity of the event horizon the spacetime 
metric for a rotating black hole in the Boyer-Lindquist coordinates 
approximates to the metric of Rindler spacetime in the throat region.  
Therefore, as long as we are interested in the throat region in the 
comoving frame, the statistical analysis (based on the string sigma 
model with the target space background given by the rotating black 
hole solution) becomes remarkably simpler.  In the following, we 
discuss some aspects of thermodynamics of the Rindler spacetime.  

The Rindler spacetime is spacetime of uniformly accelerating 
observer, called Rindler observer, in the Minkowski spacetime.  
The Rindler spacetime covers only a quadrant of Minkowski spacetime, 
defined by the wedge $x>|t|$, where $x$ is in the direction of 
uniform acceleration.  Since the trajectory of the Rindler observer's 
motion approaches the null rays $u\equiv t-x=0$ and $v\equiv t+x=0$ 
asymptotically as $t\to\pm\infty$, these rays act as event horizons.
The metric of the Rindler spacetime has the following form:
\begin{equation}
ds^2=\rho^2d\tau^2+d\rho^2+d\vec{x}^2_{\perp},
\label{rindler}
\end{equation}
where $\tau$ is the (Euclidean) Rindler time and $\vec{x}_{\perp}$ 
is the coordinates of the space parallel to the event horizon located 
at the surface $\rho=0$.  

After the time scale transformation followed by the Euclidean time 
coordinate transformation, the metric of the throat region 
of rotating black holes in the comoving frame of Hartle-Hawking vacuum 
becomes the metric of the Rindler spacetime.  In the following, we 
discuss the relationship between thermodynamic observerbles in the 
comoving frame and the Rindler frame, closely following 
Refs.\cite{HALrs392,HAL68,HAL75}.  Note, the rotation of particles 
(strings) which are in equilibrium with rotating black holes is not 
rigid, but have locally different angular velocities.  Therefore, there 
is no globally static coordinates for particles (strings).  
But, for the simplicity of calculation, in the following we assume that 
particles (strings) uniformly rotate with the angular velocity of the 
event horizon, or in other words the angular momenta of particles 
(strings) are zero in the comoving frame since the coordinate frame 
rotates with them.  This approximation can be justified since we are 
assuming that the contribution to the statistical entropy is from 
strings which are nearby the event horizon and, therefore, are somehow 
entangled with the event horizon.  Another setback of 
the comoving frame description of black hole entropy 
(in general, without restriction to the throat region) is that 
the fields in the region beyond the `velocity of light 
surface' have to be excluded, since the comoving observer has to 
move faster than the speed of light in this region or equivalently 
(as we discussed in section \ref{statcmf}) thermodynamic observables in this 
region are not well-defined.  This velocity of light surface approaches 
horizon as the angular momenta of black holes become large.  
For these reasons, we speculate that the comoving frame description 
of black hole entropy becomes accurate for small values of angular momenta.    
  
First, we consider the Rindler spacetime with metric given 
by Eq.(\ref{rindler}).  
The partition function for canonical ensemble of fields  
is given by Eq.(\ref{partition}) with the Hamiltonian $H_R$ being the 
generator of the $\tau$-translations, i.e. $H_R={{\partial}\over 
{\partial\tau}}$, which satisfies the commutation relation 
$[H_R,\tau]=1$.  One can apply the usual thermodynamic relations 
to obtain other thermodynamic observables of the canonical 
ensemble in the Rindler spacetime.  
In particular, the temperature in the Rindler spacetime 
(with which the canonical ensemble of fields is in thermal 
equilibrium) is always $T_R={1\over{2\pi}}$, which follows 
from the requirement of the absence of the conical singularity 
in the $(\tau,\rho)$-plane.  
The first law of thermodynamics of canonical ensemble of fields 
which are in thermal equilibrium with the Rindler spacetime, 
therefore, takes the following form:
\begin{equation}
dE_R={1\over{2\pi}}dS_R,
\label{1strindler}
\end{equation}
where $E_R$ and $S_R$ are respectively energy and entropy of fields 
in the Rindler spacetime.   

Second, for the spacetime of throat region in the comoving frame with 
metric given by Eq.(\ref{rindlermet}), the similar analysis can be 
done.  The Hamiltonian $H$ for fields is the generator 
for the $t$-translation, i.e. $H={{\partial}\over{\partial t}}$, 
which satisfies the commutation relation $[H,t]=1$.  
The temperature of this spacetime is $T={{\kappa}\over{2\pi}}$, 
where $\kappa$ is the surface gravity at the event horizon of the 
rotating black holes.  
The first law of thermodynamics of canonical ensemble of fields 
which are in thermal equilibrium with this spacetime is, therefore, 
given by
\begin{equation}
dE={{\kappa}\over{2\pi}}dS,
\label{1stcomoving}
\end{equation}
where $E$ and $S$ are respectively the energy and the entropy of 
fields in the comoving frame.  

Note, the time coordinates $\tau$ and $t$ of both the spacetimes are 
related by the rescaling $\tau=\kappa t$ (up to imaginary number $i$).  
So, the energy $E$ of fields is related to 
the Rindler energy $E_R$ as $dE_R={1\over\kappa}dE$.  
(This can be understood from the relationship $1=[E_R,\tau]=
\kappa[E_R,t]=\kappa{{\partial E_R}\over{\partial E}}$.)  
Therefore, it follows from (\ref{1stcomoving}) that 
entropy $S$ of fields in the comoving frame 
is related to the Rindler energy $E_R$ in the following way
\begin{equation}
S=2\pi E_R.
\label{entrindle}
\end{equation}

\subsection{Thermodynamics of Strings}\label{therstr}

In this section, we discuss thermodynamics of non-interacting 
string gas in perturbative string theory in flat target space 
background.  Statistical properties of superstring gas are 
obtained by calculating energy level densities and using 
equipartition.  These are already well-known, but we will discuss 
some of aspects for the sake of setting up notations and preparing 
for discussions in the later sections.  

In general, for the purpose of calculating energy level density 
of strings one introduces the generating function
\begin{equation}
\Pi(q)\equiv \sum^{\infty}_{n=1}d(n)q^n,
\label{genfuncden}
\end{equation}
where $d(n)$ is the degeneracy of states at the oscillator level 
$N=n$.  For bosonic and fermionic $m$-oscillators, the corresponding 
generating functions $\pi^b_m(q)$ and $\pi^f_m(q)$ take the 
following forms:
\begin{eqnarray}
\pi^b_m(q)&=&1+q^m+q^{2m}+\cdots={1\over{1-q^m}},
\cr
\pi^f_m(q)&=&1+q^m.
\label{singleocsil}
\end{eqnarray}
Each oscillator (labeled by $m$) contributes to the generating function 
multiplicatively.  Therefore, the generating function has the following 
structure:
\begin{equation}
\Pi(q)=\Pi_B(q)\Pi_F(q),
\label{gengenfunc}
\end{equation}
where $\Pi_B(q)$ and $\Pi_F(q)$ are, respectively, bosonic and 
fermionic contributions to the generating function, and take the 
following forms:
\begin{equation}
\Pi_B(q)=\prod^{\infty}_{m=1}(1-q^m)^{-n_b}, \ \ \ \ \ \ 
\Pi_F(q)=\prod^{\infty}_{m=1}(1+q^m)^{n_f}.
\label{bosfermgenfunc}
\end{equation}
Here, $n_b$ and $n_f$ depend on the degrees of freedom 
of bosonic and fermionic coordinates.  For heterotic strings, 
the bosonic factor $\Pi_B(q)$ has an additional multiplicative 
factor originated from states from the compactification on 
the sixteen-dimensional, self-dual, even, integral lattice.  

In order to obtain the degeneracy $d(n)$ of string states 
at the $n$-th oscillator level from the generating function 
$\Pi(q)$, one performs the following contour integral:
\begin{equation}
d(n)={1\over{2\pi i}}\int_C{{dz}\over z}{1\over{z^n}}\Pi(z).
\label{contour}
\end{equation}
By applying the saddle-point approximation of the contour integral, 
one obtains the following large-$n$ approximation for $d(n)$:
\begin{equation}
d(n)\simeq n^{-(D+1)/4}e^{\sqrt{n/\alpha^{\prime}}\beta_H}, 
\label{largnapprox}
\end{equation}
where $D$ is the dimension of spacetime in which strings live, 
i.e. $D=10$ [$D=26$] for superstrings [bosonic strings].   
Here, the parameter $\beta_H$ depends on fermionic spectrum: 
$\beta_H/\sqrt{\alpha^{\prime}}=\pi\sqrt{D-2}$ with fermions 
and $\beta_H/\sqrt{\alpha^{\prime}}=\pi\sqrt{{2\over3}(D-2)}$ 
without fermions.  Then, the density of states $\rho(m)$ in 
mass $m$ has the following general form that resembles the 
density of hadron level obtained from, for example, dual models:
\begin{equation}
\rho(m)\simeq md(m)\simeq cm^{-a}{\rm exp}(bm),
\label{denstatemss}
\end{equation}
where for heterotic strings $a=10$ and $b=(2+\sqrt{2})\pi
\sqrt{\alpha^{\prime}}$, for type-I superstrings 
$a={9\over 2}$ and $b=\pi\sqrt{8\alpha^{\prime}}$, and 
for closed superstrings $a=10$ and $b=\pi\sqrt{8\alpha^{\prime}}$.  

From the density of states $\rho(m)$, one obtains the partition 
function $Z(V,T)$ for canonical ensemble of (relativistic) superstrings 
enclosed in a box of volume $V$ at a temperature $T$ by following the 
procedure developed in Refs.\cite{HUAw25,HAG3,HAG5}:
\begin{eqnarray}
{\rm ln}\,Z&=&{V\over{(2\pi)^9}}\int dm\,\rho(m)\int d^9k\,
{\rm ln}{{1+e^{-\beta\sqrt{k^2+m^2}}}\over{1-e^{-\beta\sqrt{k^2+m^2}}}}
\cr
&\simeq&V\sum^{\infty}_{n=0}\left({1\over{2n+1}}\right)^5
\int\rho(m)m^{-a+5}K_5[(2n+1)\beta m]e^{bm}dm
\cr
&\simeq&\left({{TT_0}\over{T_0-T}}\right)^{-a+11/2}
\Gamma\left[-a+{{11}\over 2},\eta\left({{T_0-T}\over{TT_0}}\right)\right],
\label{strpartition}
\end{eqnarray}
where $\beta=1/T$, $K_5$ is the modified Bessel function with 
the asymptotic expansion of the form 
$K_5(z)\simeq\left({{\pi}\over{2z}}\right)^{1/2}e^{-z}\left(1+{{15}
\over{8z}}+\cdots\right)$, $\Gamma(a,x)$ is the incomplete gamma function, 
and $T_0=1/b$.  Note, the partition function $Z$ diverges 
for $T>T_0$, implying that $T_0$ is the maximum temperature for 
thermodynamic equilibrium, i.e. the Hagedorn temperature \cite{HAG3}.  
From the above expression for the partition function, one can 
calculate the thermodynamic observables: 
$P=T\partial{\rm ln}Z/\partial V$, $C_V=d\langle E\rangle/dT$, and 
$\langle E\rangle=T^2\partial{\rm ln}Z/\partial T$, etc..  

The statistical entropy $S_{stat}$ of gas of strings is defined by the 
logarithm of the degeneracy of string states at oscillator levels 
$(N_R,N_L)$:
\begin{equation}
S_{stat}={\rm ln}\,d(N_R,N_L)\simeq  
2\pi\left(\sqrt{{c_R}\over 6}\sqrt{N_R}+\sqrt{{c_L}\over 6}
\sqrt{N_L}\right)=S_R+S_L,
\label{statentstr}
\end{equation}
where $c_{R,L}=n^b_{R,L}+{1\over 2}n^f_{R,L}$ are the central charges, 
which are determined by the bosonic $n^b_{R,L}$ and the fermionic 
$n^f_{R,L}$ degrees of freedom in the right-moving and the left-moving 
sectors, and $N_R$ [$N_L$] is the right [left] moving oscillator level.  
Here, $S_R$ and $S_L$ are respectively the entropies of the left-movers 
and the right-movers.  Namely, since we assume that string gas is 
non-interacting, the total entropy $S_{stat}$ is expressed as 
the sum of contributions of two mutually non-interacting sectors, i.e. 
the right-moving sector and the right-moving sector.

\subsection{String Thermodynamics in Rindler Spacetime}\label{strthrinsp}

We discuss the statistical mechanics of strings in the Rindler spacetime 
or the comoving frame.  In general, statistical mechanics of strings 
in rotating black hole background is involved due to non-trivially 
complicated target space manifold for string sigma model.  
However, in the picture of black hole entropy, proposed by 
't Hooft \cite{HOO256} and generalized by Susskind \cite{SUS45,SUSu50}
to the case of string theories, i.e. entropy of black holes is 
nothing but entropy of strings {\it nearby horizon} which are in 
thermal equilibrium with black holes, the analysis of string 
sigma model gets simplified.  
In the throat region of the comoving frame that rotates with 
rotating black holes, the spacetime gets approximated to the Rindler 
spacetime.  Therefore, one can apply the relatively well-understood 
statistical mechanics of perturbative strings in the target space 
background of the Rindler spacetime for studying statistical 
entropy of rotating black holes.  

In the past, analysis of statistical mechanics of non-extreme black 
holes in string theories had a setback due to mismatch in dependence on 
mass of the densities of states of black holes and the string level density.  
Namely, whereas the number of states in non-extreme black holes grows with 
the ADM mass $M_{ADM}$ like $\sim e^{M^2_{ADM}}$, the string state level 
density grows with the string state mass $M_{str}$ as $\sim e^{M_{str}}$.  
So, if one takes this fact literally, then for large enough mass 
the quantum states of non-extreme black holes are much denser than 
those of a perturbative strings with the same mass.  
In Ref.\cite{SUS45}, Susskind proposed to identify the square of the mass of 
non-extreme black hole with the mass of string states, i.e. $M^2_{ADM}=
M_{str}$, in order to remedy the mismatch.  
Susskind justified such an identification by postulating that 
this is due to some unknown quantum correction of black hole mass.  
In the picture of Rindler spacetime description \cite{HALrs392}, 
such mismatch can be understood as being originated from the blueshift of 
the energy of the string oscillations in the gravitational field of black 
holes
\footnote{In the picture of ``correspondence principle" \cite{HORp46}, 
which will be discussed in the later section,  the mass of 
black holes $M_{ADM}\sim r_H/{2G_N}$ cannot always be equal to 
the mass of string states $M_{str}\sim\sqrt{N/\alpha^{\prime}}$, 
for any values of string coupling $g_s$.  Namely, since 
the gravitational constant $G_N$ depends on the string coupling 
$g_s$ and $\alpha^{\prime}$ as $G_N\sim g^2_s\alpha^{\prime}$, 
the ratio of the two masses $M_{str}/M_{BPS}$ is a function of 
string coupling $g_s$, and becomes one only at the particular value 
of string coupling $g_s$.  At this critical value of string 
coupling $g_s$, the size of horizon $r_H$ becomes of the order of 
string scale $l_s=\sqrt{\alpha^{\prime}}$, i.e. $r_H\sim l_s$, 
meaning that strings begin to form black holes due to strong 
gravitational field.  And at this critical point, the density of 
quantum states of black holes agrees with that of perturbative 
string states.}.
This blueshift of the string oscillation energy causes the rescaling of 
string tension (redshieft of the string tension) and string oscillation 
levels from its free string value in the Minkowski spacetime.  
Taking this effect into account, it is shown \cite{HAL68,HAL75} that 
the Bekenstein-Hawking entropy of non-extreme, non-rotating black holes 
can be reproduced by counting degeneracy of perturbative string states, 
only.  It is one of the purposes of this paper to generalize this picture 
to the case of rotating black holes.  

In the Rindler spacetime description of black holes proposed in 
Ref.\cite{HALrs392}, black hole 
configuration is divided into two sectors: the first sector carrying 
(perturbative) NS-NS electric charges and the second sector carrying 
non-perturbative charges, i.e. NS-NS magnetic charges and R-R charges.  
In the weak string coupling limit, NS-NS electric charges are carried 
by perturbative string states and non-perturbative charges are carried 
by black holes, which act as {\it backgrounds} in which strings oscillate.    
In this description, back-reaction of massive string states on the 
gravitational field of non-perturbative-charge-carrying black holes is 
neglected.  (Therefore, for example the mass of the whole black hole 
configuration is just a sum of the mass of perturbative string state and 
the ADM mass of non-perturbative-charge-carrying background black hole.)  
Note also that even in the weak string coupling limit the carrier of 
non-perturbative charges remain as black holes, rather than becoming, 
for example, the bound state of $D$-branes.  
However, due to the non-trivial background geometry the string tension 
and string oscillator levels get scaled relative to the free string value.  

Although the Rindler spacetime is a flat spacetime, the quantization of 
strings is non-trivial due to the event horizon and the fact that 
strings are extended objects.  The quantum theory of strings in the 
Rindler spacetime (or the uniformly accelerating frame in the Minkowski 
spacetime) is similar to the case of point-like particles except some 
modifications due to the fact that strings are extended objects.   
In the following, we discuss some of aspects of string 
theories in the Rindler spacetime in relation to statistical 
entropy of rotating black holes.  

The coordinates $X^A$ of strings in a uniformly accelerating frame of 
the flat spacetime is related to the coordinates $\hat{X}^A$ of the inertial 
frame through the following transformations:
\begin{eqnarray}
\hat{X}^1-\hat{X}^0&=&e^{\alpha(X^1-X^0)}, 
\cr
\hat{X}^1+\hat{X}^0&=&e^{\alpha(X^1+X^0)},
\cr
\hat{X}^i&=&X^i, \ \ \ \ 2\leq i\leq D-1,
\label{rindinertran}
\end{eqnarray}
where a constant $\alpha$ defines the proper acceleration of the 
Rindler observers and $D$ is the spacetime dimensions of the theory.  
Therefore, in a uniformly accelerating frame the target space metric 
$G_{AB}(X)$ takes the following form
\begin{equation}
G_{AB}(X)dX^AdX^B=\alpha^2e^{2\alpha X^1}\left[(dX^1)^2-(dX^0)^2\right]
+(dX^i)^2.
\label{trftmetaccel}
\end{equation}
Defining the light-cone variables
\begin{equation}
U\equiv X^1-X^0, \ \ \ V\equiv X^1+X^0,\ \ \ x_{\pm}=\sigma\pm\tau,
\label{lightconecrds}
\end{equation}
where $(\tau,\sigma)$ are the worldsheet coordinates, one obtains 
the following sigma-model action in the uniformly accelerating frame
\begin{eqnarray}
S_{string}&=&{1\over{2\pi\alpha^{\prime}}}\int d\sigma d\tau\sqrt{g}
g^{\alpha\beta}G_{AB}(X)\partial_{\alpha}X^A\partial_{\beta}X^B
\cr
&=&{1\over{2\pi\alpha^{\prime}}}\int d\sigma d\tau\left[
\alpha^2e^{\alpha(U+V)}\partial_{\beta}U\partial^{\beta}V+
\partial_{\beta}X^i\partial^{\beta}X^i\right],
\label{sigrindler}
\end{eqnarray}
where the worldsheet metric $g_{\alpha\beta}$ is in the conformal gauge, 
i.e. $g=\rho(\sigma,\tau){\rm diag}(-1,1)$.  The bosonic coordinates 
$X^A(\sigma,\tau)$ satisfy the usual string boundary conditions, e.g. 
$X^A(\sigma+2\pi,\tau)=X^A(\sigma,\tau)$ for closed strings.  

One can understand the coordinate transformations (\ref{rindinertran}) 
in terms of symmetries of the string sigma model and the basic 
properties of quantum field theory in the following way.   

Due to the invariance of the string sigma model action under the 
worldsheet-coordinate reparametrization $\xi^{\alpha}\to\xi^{\alpha}+
\epsilon^{\alpha}(\xi)$, where $\xi^{\alpha}\equiv(\tau,\sigma)$ 
are the worldsheet coordinates of strings, 
one can bring the worldsheet metric $g_{\alpha\beta}(\xi)$ into the 
conformal form mentioned above.   The conformal gauge fixed sigma-model 
action still has the world sheet coordinate reparametrization invariance of 
the following form:
\begin{equation}
x_+=f(x^{\prime}_+),\ \ \ \ \ \ 
x_-=g(x^{\prime}_-).
\label{confgagreparm}
\end{equation}

Note, in quantum field theories it is well-known that Fock spaces 
built from the canonical states are different for different coordinate 
basis.   Namely, the vacua defined by positive frequency states for a 
given time coordinate are not vacuum states of another coordinate 
basis with different time-coordinate.   Specializing to the case of 
string theories, the positive frequency modes defined in the 
worldsheet reparametrization transformed basis  
$\xi^{\prime\,\alpha}=(\tau^{\prime},\sigma^{\prime})$ (Cf. see 
Eq.(\ref{confgagreparm})) are not positive frequency modes with 
respect to the target space time-coordinate $X^0$.  In order to 
construct well-defined positive frequency modes (which are required 
for defining particle states in a given reference frame) in the new 
worldsheet coordinates $\xi^{\prime}=(\tau^{\prime},\sigma^{\prime})$, 
one has to perform the following target space coordinate transformations:
\begin{eqnarray}
X_1-X_0+\epsilon&=&f(X^{\prime}_1-X^{\prime}_0),
\cr
X_1+X_0+\epsilon&=&g(X^{\prime}_1+X^{\prime}_0),
\cr
X^i&=&X^{\prime\,i},\ \ \ \ 2\leq i\leq D-1.
\label{trgtcrdtran}
\end{eqnarray}
New frame parameterized by the coordinates $X^{\prime\,A}$ 
corresponds to an accelerated reference frame with acceleration 
$a=[f^{\prime}g^{\prime}]^{-1/2}\partial_{X^{\prime}_1}[{\rm ln}
(f^{\prime}g^{\prime})]$.  Now, the positive frequency modes with 
respect to the worldsheet time coordinate $\tau^{\prime}$ 
correspond to positive frequency modes with respect to 
the new target space time-coordinate $X^{\prime\,0}$. 
Here, a constant $\epsilon$ is related to the ultra-violet cut-off 
$H$ (through $\epsilon\simeq e^{-\alpha H}$) 
on the negative Rindler coordinate $X^1$ (namely, $X^1\geq -H$), 
which regularizes the divergence in the free energy and entropy of 
quantum fields due to the existence of the Rindler horizon.  
This regulator shifts the Rindler horizon at $x^1=|x^0|$ to 
the hyperbola defined by $(x^1)^2-(x^0)^2=e^{-2\alpha H}\simeq
\epsilon^2$.  

In particular, the particular case with $f=g={\rm exp}(\alpha U^{\prime})$ 
corresponds to the boost coordinate transformations (\ref{rindinertran}) 
between an accelerating frame and the inertial frame, with the 
ultra-violet cut-off taken into account.  (Note, in the notation 
of the transformations (\ref{confgagreparm}) and (\ref{trgtcrdtran}), 
the inertial frame coordinates are $\xi^{\alpha}=(\tau,\sigma)$ and $X^A$ 
without primes, and the accelerating frame coordinates are $\xi^{\prime\,
\alpha}=(\tau^{\prime},\sigma^{\prime})$ and $X^{\prime\,A}$ with primes.)  
Therefore, one can think of the boost transformation (\ref{rindinertran}) 
in the target spacetime as a subset of worldsheet coordinate 
reparametrization symmetry.   As a consequence, the string state spectrum 
of strings in the Rindler spacetime (or comoving frame) has to be in 
one-to-one correspondence with that of strings in the Minkowski spacetime 
(or inertial frame).  Namely, the mass spectra and level densities in both 
the frames have to have the same structures since they are related by one of 
symmetries of string sigma-model, i.e. worldsheet reparametrization 
invariance.  Equivalently, the boost transformations 
(\ref{rindinertran}) in the target spacetime have to be accompanied by 
the appropriate worldsheet reparametrization transformation, so that 
the positive frequency modes have well-defined meaning.  To put it 
another way, in order to define the light-cone gauge, in which the 
target space time-coordinate is proportional to the worldsheet 
time-coordinate with the proportionality constant given by the zero 
mode of the string center-of-mass frame momentum, one has to 
simultaneously perform worldsheet coordinate transformations.  

In terms of new worldsheet coordinates $\xi^{\prime\,\alpha}=
(\tau^{\prime},\sigma^{\prime})$, the periodicity of the target space 
coordinates of closed strings is modified to $\Pi_{\epsilon}=
{1\over\alpha}{\rm ln}\left({{2\pi}\over\epsilon}+1\right)$.  
So, in particular one can see that the ultra-violet cut-off was needed 
in order to insure the finite period for closed strings.  
The frequencies of the Rindler frame basis modes differ from those 
of the inertial frame modes by a factor ${{2\pi}\over{\Pi_{\epsilon}}}$.  
Also, the momentum zero modes of the bosonic coordinate expansions in 
terms of the Rindler frame orthonormal basis are differ from those in 
terms of the inertial frame orthonormal basis by the factor of 
${{2\pi}\over{\Pi_{\epsilon}}}$.  
As a consequence, the string oscillation levels in the Rindler frame 
get rescaled relative to their Minkowski spacetime values.  
This is related to the fact that the string length in the Rindler frame is 
different from its length in the inertial frame.  This is a reminiscence   
of the rescaling of the effective length of open strings 
when $D$-branes are present \cite{MALs475} in the $D$-brane picture 
of black holes.  

The creation operators and the annihilation operators of the inertial frame 
and the Rindler frame are related by a Bogoliubov transformation.  
Just as in the case of point-like particle quantum field theories, 
the expectation value of the Rindler frame number operator with respect 
to the inertial frame vacuum state has characteristic of Planckian spectrum 
at temperature $T={\alpha\over{2\pi}}$, with an additional correction term 
of the Rayleigh-Jeans type due to the fact that string has a finite length.  
In other words, the uniformly accelerating observer (the Rindler 
observer) in the Minkowski vacuum state will detect thermal radiation 
of strings at temperature $T={\alpha\over{2\pi}}$.  

The mass formulae of string states in the inertial frame and the 
Rindler frame have different forms, but have the same eigenvalues.  
This is in accordance with the expectation that the mass of 
string states should not change since the sum of the mass of 
perturbative string states and the ADM mass of background 
black hole has to remain equal to the ADM mass of the whole 
black hole configuration \cite{HALrs392,HAL68,HAL75}.  
Furthermore, the mass spectra as measured in the accelerating 
frame and in the inertial frame have the same structure.  
Therefore, the thermodynamic relations of strings in the Rindler 
frame (i.e. the throat region in the comoving frame of rotating 
black holes) have same form as those in the inertial frame 
(i.e. the Minkowski spacetime), except that the string 
oscillation levels and string tension are rescaled.  

In the following, we discuss some of thermodynamic relations of string gas 
in the Rindler frame.  In general, we believe that the argument on string 
thermodynamics is slightly different from the one in section \ref{thrindsp}, 
which we think applies rather to point-like particle field theory, as 
oppose to the formalism followed in Refs.\cite{HALrs392,HAL68,HAL75}.  
Namely, unlike particles strings have the left-moving and the right-moving 
degrees of freedoms.  Due to the main assumption of this paper that the 
gas of strings is non-interacting, the left-moving and the right-moving 
sectors of the gas of strings form two separate thermodynamic systems  
with different equilibrium temperatures $T_L$ and $T_R$, respectively.  
The total entropy $S$ of the whole system is therefore the sum of the 
entropy $S_L$ of the left-moving sector and the entropy $S_R$ of the 
right-moving sector, i.e. $S=S_L+S_R$.  Also, the total energy $E$ is 
splited into the left-moving and the right-moving pieces, i.e. 
$E=E_L+E_R$.   Therefore, the first law of thermodynamics of string gas 
has to be considered separately for the left-movers and the right-movers 
in the following way:
\begin{equation}
dS_L={1\over{T_L}}dE_L,\ \ \ \ \ \ \ \ 
dS_R={1\over{T_R}}dE_R.
\label{lrstrfstlaw}
\end{equation}
The temperatures $T_L$ and $T_R$ of the left-movers and the right-movers 
are related to the Hagedorn temperature $T$ at which weakly coupled 
strings radiate in the following way:
\begin{equation}
{1\over T}={{dS}\over{dE}}={{d(S_L+S_R)}\over{dE}}=
2\left({{dS_L}\over{dE_L}}+{{dS_R}\over{dE_R}}\right)=
2\left({1\over{T_L}}+{1\over{T_R}}\right),
\label{halglrtmprel}
\end{equation}
where we made use of the fact that $\delta E=\delta E_L+\delta E_R=2\delta 
E_L=2\delta E_R$ when the total momentum is fixed (i.e. $\delta P=\delta 
E_L-\delta E_R=0$) \cite{DEAs55}.

As it is pointed out in section \ref{thrindsp}, due to the scaling of the 
time-coordinate $t$ to the Rindler space time-coordinate $\tau$ by the 
factor of the surface gravity $\kappa=2\pi T_H$ at the event horizon 
of the background black hole, i.e. $\tau=\kappa t$, the energy of the 
left-moving (the right-moving) strings in the Rindler frame 
$E_{Rindler\,L,R}$ is related to the energy of strings in the comoving 
frame $E_{L,R}$ as $dE_{Rindler\,L,R}={1\over\kappa}dE_{L,R}$.  
So, from the first laws of thermodynamics of the left-moving and the 
right-moving sectors of strings in Eq.(\ref{lrstrfstlaw}), one has 
the following relations of the total energies of the gas of 
the left-moving and the right-moving strings in the Rindler frame 
$E_{Rindler\,L,R}$ to the entropies of the left-moving 
and the right-moving strings $S_{L,R}$:
\begin{equation}
dS_L={\kappa\over{T_L}}dE_{Rindler\,L}, \ \ \ \ \ 
dS_R={\kappa\over{T_R}}dE_{Rindler\,R}.  
\label{rindlentrel}
\end{equation}

\section{Rotating Black Holes in Heterotic String on Tori}\label{bhsol} 

In this section, we summarize properties of the general class of 
charged, rotating black hole solutions in heterotic string on tori 
which are constructed in Refs.\cite{CVEy476,CVEy54,CVEy477}.

\subsection{Electrically Charged Rotating Black Holes in Toroidally 
Compactified Heterotic Strings in $D$-Dimensions}\label{ddbhsol}

We summarize the generating solution for general 
rotating black holes which are electrically charged under the $U(1)$ 
gauge fields in heterotic string on tori, constructed in 
Ref.\cite{CVEy477}.   Such solutions 
are parameterized by the non-extremality parameter $m$, the angular 
momentum parameters $l_i$ ($i=1,...,[{{D-1}\over 2}]$), and 
two electric charges $Q^{(1)}_1$ and $Q^{(2)}_1$, which correspond to the 
electric charges of a Kaluza-Klein and a two-form $U(1)$ 
gauge fields that are associated with the same compactification 
circle of the tori.  Here, $m$ and $l_i$ are, respectively, related 
to the ADM mass and the angular momenta per unit ADM mass of 
the $D$-dimensional Kerr solution \cite{MYEp172}.  

The generating solutions are constructed by imposing $SO(1,1)$ boost 
transformations in the $O(11-D,27-D)$ $T$-duality symmetry group of 
the $(D-1)$-dimensional action.  The ADM mass $M$, the angular momenta 
$J_i$ and the electric charges of the generating solution in $D$ 
dimensions are given in terms of the parameters 
$\delta_1$ and $\delta_2$ of the $SO(1,1)$ boost transformations, and 
the parameters $m$ and $l_i$ by 
\begin{eqnarray}
M_{ADM}&=&{{\Omega_{D-2}m}\over{8\pi G_D}}[(D-3)(\cosh^2\delta_1+
\cosh^2\delta_2)-(D-4)], 
\cr
J_i&=&{{\Omega_{D-2}}\over{4\pi G_D}}ml_i\cosh\delta_1\cosh\delta_2,
\cr
Q^{(1)}_1&=&{{\Omega_{D-2}}\over{8\pi G_D}}(D-3)m\cosh\delta_1
\sinh\delta_1, 
\cr
Q^{(2)}_1&=&{{\Omega_{D-2}}\over{8\pi G_D}}(D-3)m\cosh\delta_2
\sinh\delta_2, 
\label{gensolpar}
\end{eqnarray}
where $\Omega_{D-2}\equiv{{2\pi^{{D-1}\over 2}}\over{\Gamma({{D-1}
\over 2})}}$ is the area of a unit $(D-2)$-sphere and $G_D$ is the 
$D$-dimensional gravitational constant.  Here, the $D$-dimensional 
gravitational constant $G_D$ is defined in terms of the ten-dimensional 
gravitational constant $G_{10}=8\pi^6g^2_s\alpha^{\prime 4}$ as 
$G_D=G_{10}/V_{10-D}$, where $V_{10-D}$ is the volume of the 
$(10-D)$-dimensional internal space.  

The Bekenstein-Hawking entropy $S_{BH}$ is determined by the following 
event horizon area $A_D$ through the relation $S_{BH}={{A_D}\over
{4G_D}}$:
\begin{equation}
A_D=2mr_H\Omega_{D-2}\cosh\delta_1\cosh\delta_2,
\label{surfarea}
\end{equation}
where $r_H$ is the (outer) event horizon determined by the equation
\begin{equation}
[\prod^{[{{D-1}\over 2}]}_{i=1}(r^2+l^2_i)-2N]_{r=r_H}=0.
\label{defhorizon}
\end{equation}
Here, $N$ is defined as $mr$ [$mr^2$] for even [odd] spacetime 
dimensions $D$.  

The Hawking temperature $T_H={{\kappa}\over{2\pi}}$ is determined 
by the following surface gravity $\kappa$ at the (outer) event horizon:
\begin{equation}
\kappa=\left.{1\over{\cosh\delta_1\cosh\delta_2}}
{{\partial_r(\Pi-2N)}\over{4N}}\right|_{r=r_H},
\label{surgravhor}
\end{equation}
where $\Pi\equiv\prod^{[{{D-1}\over 2}]}_{i=1}(r^2+l^2_i)$.  

The following angular velocity $\Omega_{H\,i}$ ($i=1,...,[{{D-1}\over 
2}]$) of the event horizon is defined by the condition that 
the Killing vector $\xi\equiv\partial/\partial t+\Omega_{H\,i}
\partial/\partial\phi_i$ is null on the event horizon, i.e. 
$\left.\xi^{\mu}\xi^{\nu}g_{\mu\nu}\right|_{r=r_H}=0$:
\begin{equation}
\Omega_{H\,i}={1\over{\cosh\delta_1\cosh\delta_2}}
{{l_i}\over{r^2_H+l^2_i}}. 
\label{angvel}
\end{equation}

\subsection{Dyonic Rotating Black Hole in Heterotic String on 
a Six-Torus}\label{4dbhsol}

Rotating black hole solution in heterotic string on a six-torus 
constructed in Ref.\cite{CVEy54} is parameterized by the non-extremality 
parameter $m$, the angular momentum $J$, a Kaluza-Klein 
and a two-form electric charges $Q_1$ and $Q_2$ associated with 
the same compactification direction, and a Kaluza-Klein 
and a two-form magnetic charges $P_1$ and $P_2$ associated with 
the same compactification direction but different compactification 
direction from that of the electric charges.  

In terms of the non-extremality parameter $m$, the rotational parameter 
$l$, and the boost parameters $\delta_{e1}$, $\delta_{e2}$, $\delta_{m1}$ 
and $\delta_{m2}$ of the $SO(1,1)$ boost transformations in the $O(8,24)$ 
$U$-duality symmetry group of the heterotic string on a seven-torus, 
the ADM mass $M_{ADM}$, the angular momentum $J$, and electric and 
magnetic charges $Q_1$, $Q_2$, $P_1$ and $P_2$ are given by
\begin{eqnarray}
M_{ADM}&=&2m(\cosh 2\delta_{e1}+\cosh 2\delta_{e2}+\cosh 2\delta_{m1}+
\cosh 2\delta_{m2}),
\cr
J&=&8lm(\cosh\delta_{e1}\cosh\delta_{e2}\cosh\delta_{m1}\cosh\delta_{m2}
-\sinh\delta_{e1}\sinh\delta_{e2}\sinh\delta_{m1}\sinh\delta_{m2}),
\cr
Q_1&=&2m\sinh 2\delta_{e1},\ \ \ \ \ \ \ \ \ \ \ \, Q_2=2m\sinh 2\delta_{e2},
\cr
P_1&=&2m\sinh 2\delta_{m1},\ \ \ \ \ \ \ \ \ \ \ P_2=2m\sinh 2\delta_{m2}.
\label{4dimsolpar}
\end{eqnarray}

The Bekenstein-Hawking entropy $S_{BH}={1\over{4G_N}}A$ is determined by 
the surface area $A=\left.\int d\theta d\phi\sqrt{g_{\theta\theta}
g_{\phi\phi}}\right|_{r=r_H}$ of the (outer) event horizon at $r=r_{H}=
m+\sqrt{m^2-l^2}$ as follows:
\begin{eqnarray}
S&=&16\pi\left[m^2\left(\prod^4_{i=1}\cosh\delta_i+\prod^4_{i=1}
\sinh\delta_i\right)+m\sqrt{m^2-l^2}\left(\prod^4_{i=1}\cosh\delta_i
-\prod^4_{i=1}\sinh\delta_i\right)\right]
\cr
&=&16\pi\left[m^2\left(\prod^4_{i=1}\cosh\delta_i+\prod^4_{i=1}
\sinh\delta_i\right)+\sqrt{m^4\left(\prod^4_{i=1}\cosh\delta_i
-\prod^4_{i=1}\sinh\delta_i\right)^2-(J/8)^2}\right],
\label{4dbhent}
\end{eqnarray}
where $\delta_{1,2,3,4}\equiv\delta_{e1,e2,m1,m2}$.  

\subsection{General Rotating Black Hole Solution in Heterotic 
String on a Five-Torus}\label{5dbhsol}

The generating solution for the general black hole solutions in 
heterotic string on a five-torus constructed in Ref.\cite{CVEy476} 
is parameterized by the non-extremality parameter $m$, two angular 
momenta $J_1$ and $J_2$, a Kaluza-Klein and the two-form electric 
charges $Q_1$ and $Q_2$ associated with the same compactification 
direction, and an electric charge $Q$ associated with the Hodge dual 
of the field strength of a two-form field in the NS-NS sector.  

In terms of the non-extremality parameter $m$, the rotational parameters 
$l_1$ and $l_2$, and parameters $\delta_1$, $\delta_2$ and $\delta$ 
of the $SO(1,1)$ boost transformations in the $O(8,24)$ $U$-duality 
group of heterotic string on a seven-torus, the ADM mass $M_{ADM}$, 
angular momenta $J_1$ and $J_2$, and electric charges $Q_1$, $Q_2$ 
and $Q$ are given by
\begin{eqnarray}
M_{ADM}&=&m(\cosh 2\delta_1+\cosh 2\delta_2+\cosh 2\delta),
\cr
J_1&=&4m(l_1\cosh\delta_1\cosh\delta_2\cosh\delta-
l_2\sinh\delta_1\sinh\delta_2\sinh\delta),
\cr
J_2&=&4m(l_2\cosh\delta_1\cosh\delta_2\cosh\delta-
l_1\sinh\delta_1\sinh\delta_2\sinh\delta),
\cr
Q_1&=&m\sinh 2\delta_1, \ \ \ \ 
Q_2=m\sinh 2\delta_2, \ \ \ \ Q=m\sinh 2\delta.
\label{5dbhpar}
\end{eqnarray}

The Bekenstein-Hawking entropy $S_{BH}={1\over{4G_N}}A$ is 
determined by the surface area $A=\left.\int d\theta d\phi_1d\phi_2
\sqrt{g_{\theta\theta}(g_{\phi_1\phi_1}g_{\phi_2\phi_2}-
g^2_{\phi_1\phi_2})}\right|_{r=r_H}$ of the (outer) event horizon 
located at $r=r_H=\left[m-{1\over 2}l^2_1-{1\over 2}l^2_2+{1\over 2}
\sqrt{\{2m-(l_1+l_2)^2\}\{2m-(l_1-l_2)^2\}}\right]^{1/2}$ as follows:
\begin{eqnarray}
S&=&4\pi \left[m\sqrt{2m- (l_1-l_2)^2}\left(\prod^3_{i=1}\cosh\delta_i 
+\prod^3_{i=1}\sinh\delta_i\right)\right.\cr
& &\ \ \ \ +\left. m\sqrt{2m- (l_1+l_2)^2}\left(\prod^3_{i=1}\cosh
\delta_i-\prod^3_{i=1}\sinh\delta_i\right)\right]
\cr
&=&4\pi\left[\sqrt{2m^3(\prod^3_{i=1}\cosh\delta_i+\prod^3_{i=1}
\sinh\delta_i)^2-\textstyle{1\over 16}(J_1-J_2)^2}
\right.\cr
& &\ \ \ \ +\left.\sqrt{2m^3\left(\prod^3_{i=1}\cosh\delta_i-\prod^3_{i=1}
\sinh\delta_i\right)^2-\textstyle{1\over 16}(J_1+J_2)^2}
\right],
\label{5dbhent}
\end{eqnarray}
where $\delta_{1,2,3}\equiv \delta_{1,2},\delta$.

\section{Statistical Interpretation of Rotating Black Holes in 
Heterotic String on Tori}\label{statent}

In this section, we elaborate on statistical interpretation of 
the Bekenstein-Hawking entropies of black hole solutions 
discussed in section \ref{bhsol}.  We find at least qualitative 
agreement between the Bekenstein-Hawking entropies and 
the statistical entropies based upon $D$-brane descriptions of 
Ref.\cite{STRv379,CALm472} and the correspondence principle of 
Ref.\cite{HORp46}.  We also speculate on the generalization of 
the Rindler space description of statistical entropy to the 
case of specific rotating black holes discussed in sections 
\ref{4dbhsol} and \ref{5dbhsol}.

\subsection{$D$-Brane Picture}\label{dbrent}

In this subsection, we attempt to give the statistical interpretation  
for the Bekenstein-Hawking entropy of the non-extreme, rotating, 
five-dimensional and four-dimensional black holes discussed in 
section \ref{bhsol} by applying  $D$-brane description of black holes 
\cite{CALm472}.  It turns out that even for non-extreme, rotating 
cases, the $D$-brane description reproduces the Bekenstein-Hawking 
entropy in the limit of large number of $D$-branes.   This is expected 
from the perspective of the correspondence principle.  In the following, 
we first discuss general formalism and then we consider the five-dimensional 
black hole and the four-dimensional black hole as special cases.  

Generally, in the $D$-brane picture of black holes the statistical 
origin of the Bekenstein-Hawking entropy is attributed to the 
oscillation degrees of freedom of open strings which stretch between 
$D$-branes.  So, the statistical entropy of black holes is 
the logarithm of the asymptotic level density of open strings given in 
Eq.(\ref{statentstr}), which we shall write again here as
\begin{equation}
S_{stat}=2\pi\sqrt{c\over 6}\left[\sqrt{N_L}+\sqrt{N_R}\right],
\label{dbrstatent}
\end{equation}
where $c=n_b+{1\over 2}n_f$ is the central charge determined by the 
bosonic $n_b$ and the fermionic $n_f$ degrees of freedoms for the 
configuration under consideration.  

The oscillator levels $N_L$ and $N_R$ are determined by the 
level matching condition in terms of NS-NS electric charges and the 
non-extremality parameter.   Since in the $D$-brane description of 
black holes $D$-branes do not play any dynamical role gravitationally, 
we assume that the ADM mass of the whole black hole configuration 
is the sum of the ADM mass of black hole that carries R-R charges and 
the mass of perturbative string states, as proposed in 
Refs.\cite{HALrs392,HAL68,HAL75}.   Of course, such an identification 
has to be made at the black hole and $D$-brane transition point, by  
following the correspondence principle of Ref.\cite{HORp46}.  
Then, equating the mass of the perturbative string states with 
the ADM mass of black holes which carry NS-NS electric charges, i.e. 
the Kaluza-Klein gauge field and the two-form gauge field electric 
charges, at the transition point, one obtains the oscillator levels 
$N_L$ and $N_R$ in terms of parameters of black hole solutions.  
The details are discussed in section \ref{corent} and 
expressions for the five-dimensional and four-dimensional cases are 
given by:
\begin{eqnarray}
N_R&\approx&\alpha^{\prime}m^2\cosh^2(\delta_2-\delta_1), \ \ \ \ \ \ \,
N_L\approx\alpha^{\prime}m^2\cosh^2(\delta_2+\delta_1), \ \ \ \ \ \ \ \, 
\text{for 5-d case},
\cr
N_R&\approx&2\alpha^{\prime}m^2\cosh^2(\delta_{e2}-\delta_{e1}), \ \ \   
N_L\approx 2\alpha^{\prime}m^2\cosh^2(\delta_{e2}+\delta_{e1}), \ \ \ \ 
\text{for 4-d case},
\label{dbrlvmtch}
\end{eqnarray}
where the boost parameters $\delta_1$ and $\delta_2$ are respectively 
associated with the Kaluza-Klein $U(1)$ gauge and the two-form $U(1)$ 
gauge electric charges, and $m$ is the non-extremality parameter.  

The central charge $c$ is determined by the total number of bosonic 
and fermionic degrees of freedom within the configuration under 
consideration.  It is in general expressed in terms of (the product of) 
the number of $D$-brane(s).  We will give the expressions for $c$ in 
the following subsections when we consider specific configurations.  

The effect of non-zero angular momenta on the statistical entropy 
of black holes can be explained in terms of conformal field theory 
technique as follows.  The details are discussed for example in 
Refs.\cite{BREmpv391,VAFw431,CALm472,CVEy477}.  So, we explain only the 
main points here.  The spatial rotation group $SO(4)$, which is 
external to $D$-brane bound states, is isomorphic to the 
$SU(2)_R\times SU(2)_L$ group.  This $SU(2)_R\times SU(2)_L$ group 
can be identified as the affine symmetry of $(4,4)$ superconformal 
algebra.  The charges $(F_R,F_L)$ of the $U(1)_R\times 
U(1)_L$ subgroup in $SU(2)_R\times SU(2)_L$ group 
(which can be interpreted as the spins of string states) 
are related to the angular momenta $(J_1,J_2)$ of the rotational 
group $SO(4)$ in the following way
\begin{equation}
J_1={1\over 2}(F_L+F_R),\ \ \ \  J_2={1\over 2}(F_R-F_L).  
\label{angmomchrel}
\end{equation}
The $U(1)_{L,R}$ current $J_{L,R}$ can be bosonized as $J_{L,R}=
\sqrt{c\over 3}\partial\phi$, and the conformal state $\Phi_{F_{L,R}}$ 
which carries $U(1)_{L,R}$ charge $F_{L,R}$ is obtained by applying 
an operator ${\rm exp}\left({{iF_{L,R}\phi}\over{\sqrt{c/3}}}\right)$ 
to the state $\Phi_0$ without $U(1)_{L,R}$ charge.  As a consequence, 
the conformal dimensions $h$'s, i.e. the eigenvalues of the Virasoro 
generators $L_0$ and $\bar{L}_0$, of the two conformal fields 
$\Phi_{F_{L,R}}$ and $\Phi_0$ are related in the following way
\begin{equation}
h_{\Phi_{F_L}}=h_{\Phi_0}+{{3F^2_L}\over{2c}},\ \ \ \ 
h_{\Phi_{F_R}}=h_{\Phi_0}+{{3F^2_R}\over{2c}}.
\label{eigviragen}
\end{equation}
This implies that the total number $N_{L_0}$ [$N_{R_0}$] of 
the left-moving [the right-moving] oscillations of spinless 
string states are reduced by the amount ${{3F^2_L}\over{2c}}$ 
[${{3F^2_R}\over{2c}}$] in comparison with the total number 
$N_L$ [$N_R$] of the left-moving [the right-moving] oscillations 
of states with the specific spin $F_L$ [$F_R$]:
\begin{eqnarray}
N_L&\to&N_{L_0}=N_L-{{3F^2_L}\over{2c}},
\cr
N_R&\to&N_{R_0}=N_R-{{3F^2_R}\over{2c}}.
\label{lrmvlevel}
\end{eqnarray}
Note, the level density $d_0$ for spinless states in a given level 
$(N_L,N_R)$ differs from the level density $d(N_L,N_R)={\rm exp}
\left[2\pi\sqrt{c\over 6}\left(\sqrt{N_L}+\sqrt{N_R}\right)\right]$ 
of all the states in the level $(N_L,N_R)$  by a numerical factor, 
which can be neglected in the limit of large $(N_L,N_R)$ when 
one takes logarithm of the level density.  Therefore, the 
statistical entropy of string states with specific spins $(F_L,F_R)$ 
is given by
\begin{eqnarray}
S_{stat}&\simeq&{\rm ln}\,d_0(N_{L_0},N_{R_0})\simeq 
2\pi\sqrt{c\over 6}\left(\sqrt{N_{L_0}}+\sqrt{N_{R_0}}\right)
\cr
&=&2\pi\left(\sqrt{{c\over 6}N_L-{{F^2_L}\over 4}}+
\sqrt{{c\over 6}N_R-{{F^2_R}\over 4}}\right),
\label{rotbhstatent}
\end{eqnarray}
where $F_{R,L}=J_1\pm J_2$.  

\subsubsection{Five-Dimensional Black Hole}

We consider the five-dimensional black hole discussed in section 
\ref{5dbhsol}.  Black hole solutions in heterotic string on 
tori can be transformed into solutions in type-II string theories  
with R-R charges by applying the string-string duality between the 
heterotic string on $T^4$ and type-IIA string on $K3$, $T$-duality 
between type-IIA string and type-IIB string (as necessary), 
and the $U$-duality of type-II string theory.  Since such duality 
transformations leave the Einstein-frame metric intact, 
the Bekenstein-Hawking entropy has the same form after the duality 
transformations.  The $D$-brane embedding 
of the five-dimensional black hole that we wish to consider in this 
section is the bound state of $Q_5$ $D\,5$-branes wrapped around a 
five-torus $T^5=T^4\times S^1$ and open strings wound around the 
circle $S^1$ with their internal momentum flowing along $S^1$.  
For this configuration, the central charge of open strings is 
given by $c=6Q_5$, where $Q_5=m\sinh 2\delta_5$.  We assume that 
$Q_5$ is very large.  Then, one has $Q_5=m\sinh 2\delta_5\approx 
m\cosh^2\delta_5$.  
Applying Eq.(\ref{rotbhstatent}) with the central charge $c=6Q_5$ 
and string oscillator levels given in Eq.(\ref{dbrlvmtch}), 
one has the following expression for statistical entropy of 
non-extreme, rotating, five-dimensional black hole:
\begin{eqnarray}
S_{stat}&\simeq&2\pi\left[\sqrt{\alpha^{\prime}m^3\cosh^2\delta_5
(\cosh\delta_1\cosh\delta_2+\sinh\delta_1\sinh\delta_2)^2-{{(J_1-J_2)^2}
\over 4}}\right.
\cr
& &+\left.\sqrt{\alpha^{\prime}m^3\cosh^2\delta_5
(\cosh\delta_1\cosh\delta_2-\sinh\delta_1\sinh\delta_2)^2-{{(J_1+J_2)^2}
\over 4}}\right].
\label{statentfivbh}
\end{eqnarray}
This agrees with the Bekenstein-Hawking entropy in Eq.(\ref{5dbhent}) 
in the limit of large $Q$ (and therefore $\sinh\delta\simeq\cosh\delta$).  
The mismatch in factors in each term is due to difference in convention for 
defining $U(1)$ charges and angular momenta.

\subsubsection{Four-Dimensional Black Hole}

We discuss the statistical interpretation for the Bekenstein-Hawking 
entropy (\ref{4dbhent}) of four-dimensional black hole solution 
in section \ref{4dbhsol}.  By applying duality transformations on 
this black hole solution, one can obtain a solution in type-II theory 
with R-R charges.  In this section, we consider $D$-brane bound 
state corresponding to intersecting $Q_6$ $D\,6$-branes and $Q_2$ 
$D\,2$-branes which are respectively wrapped around a six-torus 
$T^6=T^4\times S^{\prime}_1\times S_1$ and the two-torus 
$T^2=S^{\prime}_1\times S_1$, and open strings wound around 
one of circles in the two-torus $T^2$, say $S_1$, with their 
momenta flowing along the same circle, i.e. $S_1$.  The central charge 
of this $D$-brane bound state is given by $c=6Q_2Q_6$, where in terms of 
boost parameters and the non-extremality parameter, the $D$-brane charges 
are given by $Q_2=2m\sinh 2\delta_{D2}$ and $Q_6=2m\sinh 2\delta_{D6}$.  
For the four-dimensional black hole, 
the $U(1)_L$ charge $F_L$ is zero (therefore, $J:=J_1=J_2$), since 
only the right-moving supersymmetry survives (corresponding to the $(4,0)$ 
superconformal theory).   We assume that $Q_2$ and $Q_5$ are very 
large, so that $Q_5=2m\sinh 2\delta_{D5}\approx 2m\cosh^2\delta_{D5}$ 
and $Q_2=2m\sinh 2\delta_{D2}\approx 2m\cosh^2\delta_{D2}$.  
Then, the general form of statistical 
entropy (\ref{rotbhstatent}) reduces to the following form: 
\begin{eqnarray}
S_{stat}&\simeq&2\pi\left[2\sqrt{2}\sqrt{\alpha^{\prime}}m^2\cosh\delta_{D2}
\cosh\delta_{D6}(\cosh\delta_1\cosh\delta_2+\sinh\delta_1\sinh\delta_2)
\right.
\cr
& &\left.+\sqrt{8\alpha^{\prime}m^4\cosh^2\delta_{D2}\cosh^2\delta_{D6}
(\cosh\delta_1\cosh\delta_2-\sinh\delta_1\sinh\delta_2)^2-J^2}\right].
\label{bdrstentfour}
\end{eqnarray}
This agrees with the Bekenstein-Hawking entropy (\ref{4dbhent}) 
of the four-dimensional black hole solutions discussed in 
section \ref{4dbhsol} in the limit of large $P_1$ and $P_2$ 
(and therefore, $\sinh\delta_{m1}\simeq\cosh\delta_{m1}$ and 
$\sinh\delta_{m2}\simeq\cosh\delta_{m2}$).

\subsection{Correspondence Principle}\label{corent}

In this subsection, we generalize the corresponding principle of 
Ref.\cite{HORp46} to the case of rotating black holes.  We consider the 
general class of electrically charged, rotating black holes in 
heterotic string on tori discussed in section \ref{ddbhsol}, and 
we shall find that statistical entropy from correspondence principle is 
in qualitative agreement with the Bekenstein-Hawking entropy of such 
solutions.  

According to $D$-brane or fundamental string description of black 
holes, black holes are regarded as the strong string coupling limit 
of the perturbative string states or the bound states of $D$-branes.  
Namely, since the gravitational constant is proportional to the square of 
string coupling constant, when string coupling is very large the 
strong gravitational field causes gravitational collapse, 
leading to the formation of black hole.  On the other hand, for a 
small value of string coupling, spacetime approaches flat spacetime 
and the theory is described by perturbative strings or 
$D$-branes.  Therefore, at the particular value of string coupling 
there exists the transition point between black holes and perturbative 
$D$-brane or string descriptions.  It is claimed in Ref.\cite{HORp46} 
that this occurs when the curvature at the event horizon 
of a black hole is of the order of string scale 
$l_s\approx\sqrt{\alpha^{\prime}}$ or equivalently when the size 
of the event horizon is of the order of string scale, i.e. 
$r_H\sim\sqrt{\alpha^{\prime}}$.  At the transition point, the mass 
of perturbative string states can be equated with the ADM mass 
of black holes, making it possible to apply the level matching condition.  

First, we relate the macroscopic quantities that characterize black 
holes to the microscopic quantities of perturbative strings by applying 
the level matching condition.  According to the correspondence principle, 
this is possible when the size of the event horizon is of the order of 
the string length scale, i.e. $r_H\sim\sqrt{\alpha^{\prime}}$.  
Since we are considering black hole solutions with the Kaluza-Klein 
and the two-form electric charges associated with the same 
compactification direction, the corresponding Virasoro condition 
for perturbative string states is the one for string compactified on 
a circle of radius $R$:
\begin{equation}
M^2_{str}=p^2_R+{4\over{\alpha^{\prime}}}N_R
=p^2_L+{4\over{\alpha^{\prime}}}N_L,
\label{virasoro}
\end{equation}
where $M_{str}$ is the mass of string states, $p_{R,L}={{n_wR}\over
{\alpha^{\prime}}}\pm{{n_p}\over R}$ is the right (left) moving momentum 
in the direction of circle, and $N_{R,L}$ is the right (left) moving 
oscillator level.  Here, $n_w$ and $n_p$ are respectively the 
string winding number and the momentum quantum number along the direction 
of the circle.  The Kaluza-Klein electric 
charge and the two-form electric charge of black holes in string theory 
are respectively identified with the momentum and the winding modes 
of strings in the compact direction.  Therefore, the right and left 
moving momenta of strings in the compact direction are given in terms 
of parameters of black hole solutions discussed in section \ref{ddbhsol} 
by 
\begin{equation}
p_{R,L}={{\Omega_{D-2}}\over{8\pi G_D}}(D-3)m(\cosh\delta_2
\sinh\delta_2\pm\cosh\delta_1\sinh\delta_1).
\label{rlmvmomentabh}
\end{equation}
When the size of the event horizon is of the length of string scale 
(i.e. $r_H\sim\sqrt{\alpha^{\prime}}$), one can further identify  
the ADM mass of the black hole with the mass of string states, i.e. 
$M_{str}\simeq {{\Omega_{D-2}m}\over{8\pi G_D}}[(D-3)(\cosh^2\delta_1+
\cosh^2\delta_2)-(D-4)]$.  Then, from the level matching condition 
(\ref{virasoro}) with (\ref{rlmvmomentabh}) substituted one obtains the 
following expressions for the right and left moving oscillation levels 
in terms of parameters of black hole solution
\begin{eqnarray}
N_R&\approx&{{\alpha^{\prime}}\over 4}\left({{\Omega_{D-2}}\over
{8\pi G_D}}\right)^2(D-3)^2m^2\cosh^2(\delta_2-\delta_1), 
\cr 
N_L&\approx&{{\alpha^{\prime}}\over 4}\left({{\Omega_{D-2}}\over
{8\pi G_D}}\right)^2(D-3)^2m^2\cosh^2(\delta_2+\delta_1), 
\label{oscilvsbh}
\end{eqnarray} 
in the limit of large electric charges $Q_1$ and $Q_2$.  

The statistical entropy of the black hole is given by the logarithm 
of the degeneracy of string states.  From the expression for 
oscillation levels in Eq.(\ref{oscilvsbh}) one obtains the following 
form of statistical entropy:
\begin{eqnarray}
S_{stat}&=&2\pi\sqrt{c\over 6}(\sqrt{N_L}+\sqrt{N_R})
\cr
&\simeq&{{(D-3)\Omega_{D-2}\sqrt{\alpha^{\prime}}m}\over
{4G_D}}\sqrt{c\over 6}\cosh\delta_1\cosh\delta_2, 
\label{statentcp}
\end{eqnarray}
in the limit of large electric charges.  
On the other hand, at the transition point, the Bekenstein-Hawking 
entropy (\ref{surfarea}) takes the following form:
\begin{equation}
S_{BH}={{A_D}\over{4G_D}}\simeq
{{m\sqrt{\alpha^{\prime}}\Omega_{D-2}}\over{2G_D}}
\cosh\delta_1\cosh\delta_2.
\label{bhenttran}
\end{equation}
Here, $m$ is a function of angular momentum parameters $l_i$ and 
$\alpha^{\prime}$, and is determined by the equation 
$\prod^{[{{D-1}\over 2}]}_{i=1}(\alpha^{\prime}+l^2_i)-2N=0$.  
Therefore, the Bekenstein-Hawking entropy (\ref{bhenttran}) 
and the statistical entropy (\ref{statentcp}) agree up to a numerical 
factor of order one.  

\subsection{Statistical Entropy and Rindler Geometry}\label{rindstent}

As for the statistical interpretation of rotating 
black hole entropy based upon Rindler space description, which is extensively 
discussed in the previous sections, the author does not have a complete 
understanding yet.  
But if such description is a right interpretation of black hole entropy, 
we believe that the entropy of rotating black holes is nothing but the 
statistical entropy of a gas of strings which rotate with 
rotating black holes that carry (the remaining) non-perturbative charges.  
In the following subsection, we shall discuss necessary ingredients for 
understanding the Rindler spacetime description for the general class of 
four-dimensional and five-dimensional black holes discussed in section 
\ref{bhsol}.

\subsubsection{Five-Dimensional Black Hole}\label{5dstatent}

The five-dimensional black hole solution constructed in 
Ref.\cite{CVEy476} carries two electric charges $Q_1$ and $Q_2$ of a 
Kaluza-Klein $U(1)$ gauge field and a two-form $U(1)$ gauge field 
in the NS-NS sector, and one electric charge $Q$ associated with the 
Hodge-dual of the field strength of the two-form field in the 
NS-NS sector.   We interpret the statistical origin of entropy of 
this black hole solution as being due to the microscopic degrees of 
freedom of gas of perturbative strings with momentum number $Q_1$ and 
the winding number $Q_2$ which oscillate in the background of black hole 
with electric charge $Q$.  

Five-dimensional, rotating black hole with the electric charge $Q$ has 
the surface gravity at the event horizon $r=r_H$ given by
\begin{eqnarray}
\kappa&=&{1\over{\cosh\delta}}{{2r^2_H+l^2_1+l^2_2-2m}\over{2mr_H}}
\cr
&=&{1\over {m\cosh\delta}}{{\sqrt{\{2m-(l_1+l_2)^2\}\{2m-(l_1-l_2)^2\}}}
\over{\sqrt{2m-(l_1+l_2)^2}+\sqrt{2m-(l_1-l_2)^2}}}.
\label{5dbacksurfgrv}
\end{eqnarray}  
The Rindler observer will detect thermal radiation of a gas of strings 
with temperature $T={\kappa\over{2\pi}}$.  The total energy of strings 
in the comoving frame with metric given by Eq.(\ref{rindlermet}) 
is related to the Rindler energy of strings in the frame with 
metric (\ref{rindler}) by the factor of the above surface gravity $\kappa$ 
at the event horizon. 

Note, in the extreme limit the surface gravity $\kappa$ 
vanishes, leading to infinite rescaling of oscillation levels or 
infinite statistical entropy.  
This is not surprising since in general in the extreme limit observers 
in the comoving frame with angular velocity of event horizon have to 
move faster than the speed of light, i.e. $g^{\prime}_{tt}\geq 0$ for 
$r\geq r_H$, and as a consequence the statistical observerbles are not 
well-defined.  So, the statistical description in the comoving frame 
which rotates with angular velocity of the event horizon 
cannot be applied to the extreme rotating black holes.  

By applying the level matching condition, one can see that 
the right moving and the left moving oscillation levels $N_{R}$ and 
$N_{L}$ (of free strings in the Minkowski background) are given by
\begin{equation}
N_R\approx\alpha^{\prime}m^2\cosh^2(\delta_2-\delta_1), \ \ \ \ \ 
N_L\approx\alpha^{\prime}m^2\cosh^2(\delta_2+\delta_1),
\label{5dbhoscilnum}
\end{equation}
in the limit of large $Q_1$ and $Q_2$.  
Naively just by applying the prescription of Ref.\cite{HALrs392,HAL68,HAL75}, 
i.e. the statistical entropy is given by the logarithm of level density 
with string oscillation levels $N_L$ and $N_R$ rescaled by the 
surface gravity $\kappa$, one does not reproduce the Bekenstein-Hawking 
entropy.  We speculate that the left-moving and the right-moving 
oscillator levels $N_L$ and $N_R$ should be rescaled by different factors  
since the left-moving and the right-moving sectors of string gas form 
separate non-interacting thermal systems with different temperatures 
$T_L$ and $T_R$.  The following rescalings of the oscillator levels 
$N_L$ and $N_R$ would yield the correct expressions for statistical 
entropy:
\begin{eqnarray}
N_L&\to&N^{\prime}_L=N_L[2m-(l_1-l_2)^2]\cosh^2\delta,
\cr
N_R&\to&N^{\prime}_R=N_R[2m-(l_1+l_2)^2]\cosh^2\delta,
\label{scilscalfiv}
\end{eqnarray}
for the five-dimensional black holes with large $Q=m\sinh 2\delta$.  
These rescaling factors cannot be understood from the temperatures 
$T_L$ and $T_R$ of the left-movers and the right-movers alone.  
So, we believe that there are some subtle points in statistical mechanics 
of string gas in the Rindler frame that we do not have a complete 
understanding of.

\subsubsection{Four-Dimensional Black Hole}\label{4dstatent}

The four-dimensional black hole solution constructed in 
Ref.\cite{CVEy54} is charged under two electric charges $Q_1$ 
and $Q_2$ of a Kaluza-Klein $U(1)$ gauge field and a two-form 
$U(1)$ gauge field in the NS-NS sector, and under two magnetic 
charges $P_1$ and $P_2$ of a Kaluza-Klein $U(1)$ gauge field and a 
two-form $U(1)$ gauge field in the NS-NS sector.  
Just as in the five-dimensional case in section \ref{5dstatent}, 
we attribute the statistical entropy of this black hole as being due 
to a gas of perturbative strings with momentum number $Q_1$ and the 
winding number $Q_2$ which oscillate in the background of black hole 
that carries magnetic charges $P_1$ and $P_2$.  

The surface gravity at the event horizon $r=r_H$ of the 
four-dimensional, rotating black hole with magnetic charges $P_1$ and 
$P_2$ is given by
\begin{equation}
\kappa={1\over{\cosh\delta_{m1}\cosh\delta_{m2}}}{{r_H-m}\over{2mr_H}}
={1\over{2m\cosh\delta_{m1}\cosh\delta_{m2}}}{\sqrt{m^2-l^2}\over
{m+\sqrt{m^2-l^2}}}.
\label{4dbhsurfgrv}
\end{equation}
As expected, in the extreme limit ($m=l$) the analysis of this 
section cannot be applied, since $\kappa=0$.  

By applying the level matching condition, one obtains the following 
expression for the oscillator levels for free strings in the 
Minkowski background:
\begin{equation}
N_R\approx 2\alpha^{\prime}m^2\cosh^2(\delta_{e2}-\delta_{e1}), \ \ \ \ \ 
N_L\approx 2\alpha^{\prime}m^2\cosh^2(\delta_{e2}+\delta_{e1}),
\label{4dbhoscilnum}
\end{equation}
in the limit of large $Q_1$ and $Q_2$.  
We speculate that in the throat region of comoving frame the 
string oscillation levels $N_L$ and $N_R$ are rescaled in the following 
way:
\begin{eqnarray}
N_L&\to&N^{\prime}_L=4m^2\cosh^2\delta_{m1}\cosh^2\delta_{m2}\,N_L,
\cr
N_R&\to&N^{\prime}_R=4(m^2-l^2)\cosh^2\delta_{m1}\cosh^2\delta_{m2}\,N_R.
\label{scilscalfou}
\end{eqnarray}
for the four-dimensional black hole with large $P_1=2m\sinh 2\delta_{m1}$ 
and $P_2=2m\sinh 2\delta_{m2}$.

\acknowledgments
The work is supported by U.S. DOE Grant No. DOE DE-FG02-90ER40542.

\end{document}